\documentclass[a4paper, oneside, twocolumn, notitlepage, 10pt]{extarticle_ecoc}
\usepackage{ecoc}

\usepackage{amsmath,amssymb}
\usepackage{pgfplots} 
\usepackage{adjustbox}
\usepackage{pgffor}
\pgfplotsset{compat=1.16}
\usepgfplotslibrary{colorbrewer}

\usepgfplotslibrary{external}
\begin{document}
\selectlanguage{english}    


\title{220-GBd optical coherent waveform generation using temporal unitary transforms}%


\author{
    Callum Deakin, Xi Chen, Di Che
}

\maketitle                  


\begin{strip}
    \begin{author_descr}

        Nokia Bell Labs, 600 Mountain Ave, Murray Hill, NJ, USA,
        \textcolor{blue}{\uline{callum.deakin@nokia-bell-labs.com}}

    \end{author_descr}
\end{strip}

\renewcommand\footnotemark{}
\renewcommand\footnoterule{}


\begin{strip}
    \begin{ecoc_abstract}
        We use temporal unitary transforms to generate 16-QAM up to 220 GBd using only 50-GHz electrical bandwidth. The technique is theoretically lossless and can generate arbitrary optical waveforms beyond the bandwidth of the constituent modulators.  ©2026 The Author(s) 
    \end{ecoc_abstract}
\end{strip}


\section{Introduction}
\begin{figure*}[b]
   \centering
       \vspace{-0.5cm}
    \includegraphics[width=\linewidth]{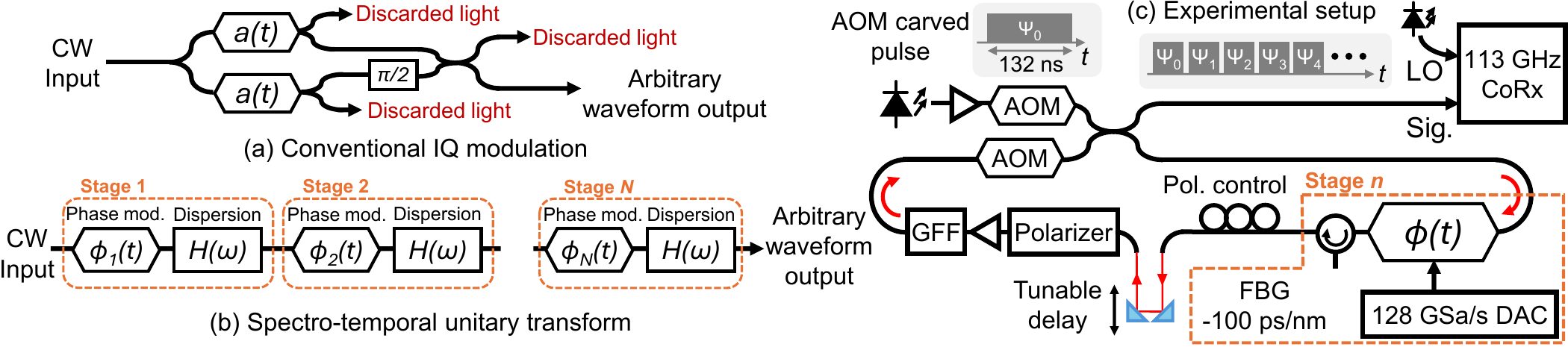}
    \vspace{-0.5cm}
    \caption{(a) Conventional IQ modulation based on amplitude modulation, $a(t)$. (b) Lossless spectro-temporal unitary transforms based on $N$ cascaded phase modulators $\phi(t)$ and dispersive elements $H(\omega)$. (c) Recirculating loop experimental setup.}
    \vspace{-0.5cm}
    \label{experiment_setup}
\end{figure*}

Optical coherent modulation typically uses two $\pi/2$ out of phase Mach-Zehnder modulators (MZMs) for each orthogonal polarization mode, as shown in Fig.~\ref{experiment_setup}(a). This allows for arbitrary manipulation of the optical waveform to maximise the spectral efficiency of the channel using advanced modulation formats. However, this scheme is fundamentally lossy since it is based on switching: excess light in low amplitude symbols is discarded, with modulation losses in modern coherent transceivers often exceeding 25~dB. Furthermore, the system bandwidth is limited by the bandwidth of the constituent MZMs, meaning that increasing the symbol rate necessarily means using higher bandwidth modulators, drivers and DACs. It is unclear whether future CMOS nodes will provide the necessary increase in bandwidth as layer interconnect parasitics begin to dominate~\cite{ryckaert2018complementary,liou2024gigahertz}, while the limited optical power in integrable tunable laser assemblies (ITLAs) will impose fundamental SNR limits due to shot noise as the optical bandwidth increases.

An alternative method of arbitrary optical waveform generation based on multiple stages of phase modulation and dispersion, as shown in Fig.~\ref{experiment_setup}(b), has been proposed~\cite{thiel2017programmable,mazur2019optical}.  This method uses only temporal and spectral phase modulations and achieves amplitude modulation by redistributing light temporally rather than switching on a symbol-by-symbol basis. This modulation is therefore lossless aside from practical limitations such as the insertion loss of the devices. In addition, the cascaded phase modulations mean that the method can generate waveforms with greater bandwidth than that of its constituent phase modulators, given sufficient modulation stages~\cite{deakinofc2025}. It has also been shown recently that this modulation scheme can be operated in real time~\cite{saxena2023performance}. Despite several theoretical and numerical investigations of this technique~\cite{saxena2023performance,deakinofc2025,deakin2025spectro}, the single published experimental demonstration so far demonstrated a maximum symbol rate of only 20~GBaud and generated low order modulation (OOK/QPSK) with poor SNR ($\approx 5$~dB)~\cite{mazur2019optical,mazur2019multi}. Furthermore, a large RF signal drive power ($\approx 20$~dBm) was used to achieve the necessary phase modulation, which is impractical for integration into a modern energy efficient coherent transceiver. 

 We present the first demonstration of this technique for coherent modulation (16-QAM) at high symbol rates (up to 220 GBd), and introduce a novel multi-objective optimization technique that allows for a substantial reduction in the required modulator driver power ($< 13$~dBm).  These results provide the first experimental evidence that this technique can be implemented in modern high baudrate coherent optical transceivers, and can generate high baudrate (220~GBd) Nyquist signalling with electrical modulation bandwidth (50~GHz, brick-wall filter) well below the Nyquist bandwidth of the signal (110~GHz). Furthermore, the lossless and universal nature of these transforms may enable their application to temporal mode transforms for ultrafast and quantum optics~\cite{ashby2020temporal,joshi2022picosecond}.

\section{RF power reduction by multi-objective optimization}
\begin{figure*}[b]
   \centering
    \input{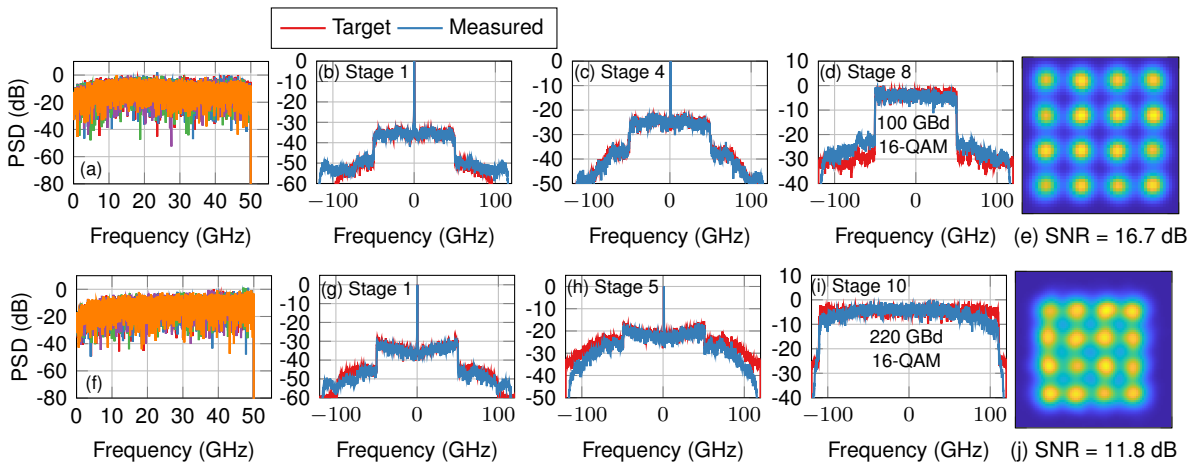}
    \caption{100 GBd 16-QAM using 8 stages: (a) Phase modulations spectra, (b-d) spectrum after 1, 4, 8 stages, (e) constellation. 220 GBd 16-QAM using 10 stages: (f) phase modulations spectra, and (g-h) spectrum after 1, 5, 10 stages, (j) constellation.}
    \label{fig:example}
\end{figure*}
The fundamental difference between this method and traditional IQ modulation is that, instead of linearly mapping the symbols (e.g. 16-QAM) to the optical domain, a series of phase modulations that results in the desired optical signal needs to be identified. Finding the set of phase modulations $\{\phi_1(t),\phi_2(t),\dots \phi_N(t)\}$ is non-trivial and analytic constructions are known for only low mode and stage counts~\cite{alvarez2025quantum,yasir2025compactifying}, while constructive decompositions result in a large number of stages~\cite{lopez2021arbitrary,girouard2025near}. Therefore, gradient descent based optimisation must be used to find the set of phase modulations that produce the desired waveform $\Psi_\textnormal{target}(t)$. In contrast to previous work, we use a multi-objective optimisation procedure to simultaneously minimise both waveform noise-to-signal ratio (NSR) and the average RF drive power. The objective function $f(mT)$ is a linear combination of the waveform NSR and average RF drive power, with a linear scalarization parameter $\alpha_{s} \in \mathbb{R}^+$ 
\begin{align}
\label{multi_objective}
    f(mT) =  \underbrace{\frac{1}{E_\textnormal{target}}\sum_{m=0}^M |\Psi_\textnormal{target}(mT) - \Psi_N(mT)|^2}_\text{waveform NSR} \nonumber \\ +\underbrace{\frac{\alpha_{s}}{M}\sum_{m=0}^M \phi_n^2({mT})}_\text{average RF drive power \nonumber} 
    \end{align}
Here $\Psi_N(t)$ is the generated waveform after $N$ stages, and $E_\textnormal{target}$ is the total energy of the waveform. All vectors represent a time sampled system with sampling period $T$ and $M$ samples. Increasing the RF driver power by increasing the scalarization parameter $\alpha_{s}$ will increase the waveform signal-to-noise ratio (SNR) and vice versa: the tradeoff between SNR and driver power can be therefore adjusted by changing the scalarization parameter $\alpha_{s}$. 

In this experiment, the target waveform $\Psi_\textnormal{target}(t)$ is a single polarisation root-raised-cosine (RRC, $\beta = 0.01$) 16-QAM signal. The L-BFGS algorithm~\cite{byrd1995limited} is used to find the set of phase modulations $\{\phi_1(t),\phi_2(t),\dots \phi_N(t)\}$ that minimise the objective function $f(mT)$. The amount of dispersion required has been shown numerically to be $\approx T_s^2$ for symbol period $T_s$~\cite{deakin2025spectro}, which is for example 19.6~ps/nm for 200~GBd at 1550~nm and achievable with on-chip dispersive devices with low loss~\cite{stern2023silicon}. Increasing dispersion beyond this does not generally improve performance~\cite{deakin2025spectro}. Although we generate only a single frame for offline processing in this experiment, an overlap-and-discard method can be used to perform this optimisation on a continuous stream of target symbols~\cite{saxena2023performance}.

\section{Experimental setup}

\begin{figure*}[t]
   \centering
    \pgfmathsetmacro{\figuresizex}{0.38}
\pgfmathsetmacro{\figuresizey}{0.23}
\tikzexternaldisable
\begin{tikzpicture}[trim axis left,trim axis right]

\begin{axis} [ylabel= RF power (dBm), 
              xlabel=Symbol rate (GBd),
              xmin=100,xmax=220,
              ymax=13,
              ymin=3,
              height=0.85*\figuresizey*\linewidth,
              grid=both,
              width=\figuresizex*\linewidth,
              cycle list/Set1,
              ylabel near ticks,
              legend columns =3,
              clip mode=individual,
              log ticks with fixed point,
              legend style={at={(0.5,1)}, anchor=south, font=\footnotesize},
              label style={font=\footnotesize},
              tick label style={font=\footnotesize},
              yticklabels={0,2,4,6,8,10,12},
              ytick={0,2,4,6,8,10,12},
              xticklabels={80,100,120,140,160,180,200,220},
              xtick={80,100,120,140,160,180,200,220}]]

\addplot+ [thick,mark=*] table [x=GBd, y=6power2, col sep=comma] {data/SymbolRate_v_SNR_2026-01-20_new.csv};
\addlegendentry{$N=6$};
\addplot+ [thick,mark=triangle*] table [x=GBd, y=8power2, col sep=comma] {data/SymbolRate_v_SNR_2026-01-20_new.csv};
\addlegendentry{$N=8$};
\addplot+ [thick,mark=square*] table [x=GBd, y=10power2, col sep=comma] {data/SymbolRate_v_SNR_2026-01-20_new.csv};
\addlegendentry{$N=10$};

\end{axis}
\end{tikzpicture}\hspace{1.2cm}\begin{tikzpicture}[trim axis left,trim axis right]

\begin{axis} [ylabel= SNR (dB), 
              xlabel=Symbol rate (GBd),
              xmin=100,xmax=220,
              ymax=19,
              ymin=10,
              height=\figuresizey*\linewidth,
              grid=both,
              width=\figuresizex*\linewidth,
              cycle list/Set1,
              ylabel near ticks,
              legend columns =2
              clip mode=individual,
              log ticks with fixed point,
              legend style={at={(0.5,0.25)}, anchor=south, font=\footnotesize},
              label style={font=\footnotesize},
              tick label style={font=\footnotesize},
              xticklabels={80,100,120,140,160,180,200,220},
              xtick={80,100,120,140,160,180,200,220}]]

\addplot+ [thick,mark=*] table [x=GBd, y=6, col sep=comma] {data/SymbolRate_v_SNR_2026-01-20_new.csv};
\addplot+ [thick,mark=triangle*] table [x=GBd, y=8, col sep=comma] {data/SymbolRate_v_SNR_2026-01-20_new.csv};
\addplot+ [thick,mark=square*] table [x=GBd, y=10, col sep=comma] {data/SymbolRate_v_SNR_2026-01-20_new.csv};

 \end{axis}
\end{tikzpicture}\hspace{1.2cm}\begin{tikzpicture}[trim axis left,trim axis right]

\begin{axis} [ylabel= NGMI, 
              xlabel=Symbol rate (GBd),
              xmin=100,xmax=220,
              ymax=1,
              ymin=0.7,
              height=\figuresizey*\linewidth,
              grid=both,
              width=\figuresizex*\linewidth,
              cycle list/Set1,
              ylabel near ticks,
              legend columns =3,
              clip mode=individual,
              log ticks with fixed point,
              legend style={at={(0.5,1.05)}, anchor=south, font=\footnotesize},
              label style={font=\footnotesize},
              tick label style={font=\footnotesize},
              xticklabels={80,100,120,140,160,180,200,220},
              xtick={80,100,120,140,160,180,200,220}]]

\addplot+ [thick,mark=*] table [x=GBd, y=6_NGMI, col sep=comma] {data/SymbolRate_v_SNR_2026-01-20_new.csv};
\addplot+ [thick,mark=triangle*] table [x=GBd, y=8_NGMI, col sep=comma] {data/SymbolRate_v_SNR_2026-01-20_new.csv};
\addplot+ [thick,mark=square*] table [x=GBd, y=10_NGMI, col sep=comma] {data/SymbolRate_v_SNR_2026-01-20_new.csv};

\draw[dashed,black,thick] (100, 0.8456) -- (220, 0.8456) node[anchor=north,pos=0.25,sloped]{{\footnotesize NGMI$_{\textnormal{th}} = 0.8456$}};   

\end{axis}
\end{tikzpicture}
    \caption{Results for 100-220~GBd 16-QAM generation for $N=6$ (red circles), $N=8$ (blue triangles), and $N=10$ (green squares). (a) Required average RF driver power, (b) measured symbol SNR and (c) NGMI, with FEC threshold NGMI$_{\textnormal{th}}$.}
    \label{fig:overall}
\end{figure*}
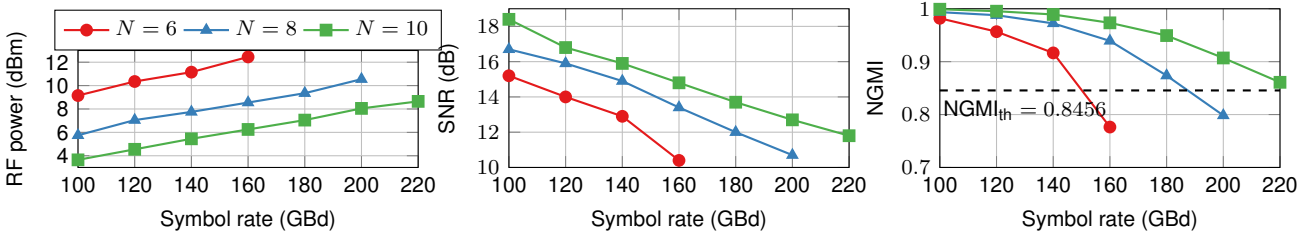

We use a recirculating loop experimental setup, shown in Fig.\ref{experiment_setup}(c), to apply the successive phase modulations since it allows programmable changes in the number of stages, $N$.  A 200~MHz acousto-optic modulator (AOM) is used to carve a 132~ns pulse from a CW laser that is launched into the recirculating loop. The laser is low linewidth (NKT Koheras X15) to simplify loop calibration but in principle a standard ITLA can be used. A bulk lithium niobate phase modulator (3~dB bandwidth $\approx 30$~GHz) and 128~GSa/s DAC modulate the light with the optimised phase modulations, followed by a fiber Bragg grating (FBG) which provides the dispersion (-100~ps/nm). A free space tunable delay line within the loop precisely sets the loop length to 144~ns ($\approx 29$~m) since misalignment between the successive phase modulations applied by the DAC on each circulation can degrade the SNR~\cite{deakin2025spectro}. An eribum doped fiber amplifier (EDFA) compensates the total per loop loss of approximately 15~dB, followed by a gain flattening filter (GFF) that minimises the gain tilt introduced by the EDFA and an AOM that terminates the recirculating light after the required number of stages. The primary contributors to the optical loss within the loop are the AOM insertion loss (4.5~dB), modulator insertion loss (3.5~dB), input/output coupler (3~dB) and GFF (3.3 dB). These losses would be eliminated in a non-loop integrated implementation (as in Fig.~\ref{experiment_setup}(b)), with expected losses per stage of $<0.5$~dB on a low loss platform such as thin film lithium niobate (TFLN)~\cite{zhu2021integrated}. Within the loop we also use a polarisation controller since the FBG and tunable delay line contain non-polarization maintaining fiber; a polarizer after the delay line ensures that the circulating light is launched back onto the correct polarisation axis of the modulator. Before applying the optimised phase modulations, we use a single stage configuration to calibrate the ampliude/phase response of both the DAC/modulator and FBG. The DAC/modulator is then compensated through digital pre-emphasis of the phase modulations, while the group delay ripple of the FBG is compensated by including the measured group delay deviation from the nominal $-100$~ps/nm in the optimization procedure. The signal exiting the loop, which is a pulse train of the optical signal after each modulation stage, is then detected by a 113 GHz bandwidth single polarisation coherent receiver with a standard ITLA as a local oscillator. The generated signal after the required number of stages is then sent to standard receiver side DSP, including a least means squared equalizer with an embedded digital phase locked loop to recover the transmitted 16-QAM symbols.

\section{Results and Discussion}

Using the aforementioned gradient descent based optimisation, we generate sets of phase modulations targeting $>25$~dB SNR. The spectra of the calculated phase modulations for the example cases \{100~GBd, $N=8$\} and \{220~GBd, $N=10$\} are shown in Fig.~\ref{fig:example}(a) and Fig.~\ref{fig:example}(f) respectively. In all cases the spectral content of the phase modulations is restricted to DC-50~GHz by applying a low pass brick wall filter in the optimisation procedure. Also shown in  Fig.~\ref{fig:example} are the measured spectra ((b-d),(g-i)) of the received optical signals (blue) at different stages, compared to the simulated target waveform (red). Notable in Fig.~\ref{fig:example}(i) is the asymmetry in the spectrum compared to the target waveform for the measured 220~GBd signal. This is caused by the residual gain tilt of the EDFA that is not compensated by the GFF. This residual gain tilt ($\approx$0.25~dB/nm) introduces a non-unitary operation into the transform that cannot be compensated by the phase modulations because it breaks the unitary symmetry of the transform, and is a major limitation of the recirculating loop setup for this experiment that causes the SNR degradation at higher baudrates, as exemplified by the constellation diagram for 220~GBd in Fig.~\ref{fig:example}(j).

Fig.~\ref{fig:overall} shows the overall results for symbol rates 100-220~GBd and $N=6,8,10$. Fig.~\ref{fig:overall}(a) shows the average RF power, which is at achievable levels ($< 13$~dBm) in all cases for current broadband integrated drivers~\cite{ozaki2025c+}. The driver power can be reduced by increasing the number of stages, $N$. Note that the RF power is calculated using the  $V_\pi = 3.1$~V @ 1~GHz and integrated S21 response of the bulk lithium niobate modulator used in this experiment: the RF power could be substantially reduced by using e.g. TFLN, which can achieve $V_\pi < 1$~V at 50~GHz bandwidth~\cite{zhu2021integrated}. The SNR and NGMI are shown in Fig.~\ref{fig:overall}(b) and Fig.~\ref{fig:overall}(c) for symbol rates 100-220~GBd. As an example, for $N=10$, the SNR decreases from 18.4~dB at 100 GBd to 11.8~dB at 220 GBd. The cause of this SNR degradation with baudrate is not fundamental and is caused by limitations of the recirculating loop setup, as well as more typical limitations of the coherent receiver (i.e. IQ imbalance, reduced ENOB, etc). Besides the previously mentioned non-unitary nature of the EDFA gain tilt, the SNR is also degraded by polarisation mode dispersion (PMD) of the FBG ($\approx 120$~fs), the limited precision of our delay line ($\approx 50$~fs), and the phase stability of the fiber loop which is 29~m long and is subject to environmental disturbances. PMD would be eliminated on an integrated platform, and the length of each stage would be reduced by several orders of magnitude to $< 1$~cm, with a corresponding improvement in phase stability/accuracy and SNR. Despite these experimental limitations, we still achieve an NGMI of 0.8610 for 220~GBd 16-QAM using $N=10$, which allows error free transmission using a rate-0.7932 forward error correction (FEC) scheme with an NGMI threshold of 0.8456~\cite{gene2020experimental}.

\section{Conclusion}

We demonstrate the modulation of error-free Nyquist 16-QAM signals up to 220~GBd using spectro-temporal unitary transforms, using only 50~GHz electrical modulation bandwidth. This method can enable low loss modulation in future high-symbol rate transceivers, without dramatically increasing the required electrical bandwidth.

\clearpage

\printbibliography

\vspace{-4mm}

\end{document}